\documentclass[useAMS,usenatbib]{mn2e}

\usepackage{psfig}

\title[A donor star signature of LMC\,X-2]{A signature of the donor star in the extra-galactic X-ray binary LMC\,X-2}
\author[R. Cornelisse et~al.]{R. Cornelisse$^{1,2}$\thanks{E-mail:
corneli@iac.es}, D. Steeghs$^{3,4}$, J. Casares$^{1}$, P.A. Charles$^{5,2}$, I.C. Shih$^{6,2}$,
\newauthor
R.I. Hynes$^{7}$, K. O'Brien$^{8}$\\
$^{1}$Instituto de Astrofisica de Canarias, Via Lactea, La Laguna E-38200, Santa Cruz de Tenerife, Spain\\
$^{2}$School of Physics and Astronomy, University of Southampton, Highfield, Southampton SO17 1BJ, UK\\
$^{3}$Harvard-Smithsonian Center for Astrophysics, 60 Garden Street, Cambridge, MA 02138, USA\\
$^{4}$Department of Physics, University of Warwick, Coventry, CV4 7AL, UK\\ 
$^{5}$South Africa Astronomical Observatory, P.O.Box 9.Observatory 7935, South Africa\\
$^{6}$Department of Physics \& Astronomy, Michigan State University, East Lansing, MI 48824, USA\\
$^{7}$Department of Physics and Astronomy, 202 Nicholson Hall, Louisiana State University, Baton Rouge, LA 70803, USA\\
$^{8}$European Southern Observatory, Casilla 19001, Santiago 19, Chile\\
}

\begin{document}

\date{Accepted  Received ; in original form }

\pagerange{\pageref{firstpage}--\pageref{lastpage}} \pubyear{2004}

\maketitle

\label{firstpage}

\begin{abstract}
  Two nights of phase-resolved medium resolution VLT spectroscopy of
  the extra-galactic low mass X-ray binary LMC\,X-2 have revealed a
  0.32$\pm$0.02 day spectroscopic period in the radial velocity curve
  of the He\,II $\lambda$4686 emission line that we interpret as the
  orbital period. However, similar to previous findings, this radial
  velocity curve shows a longer term variation that is most likely due
  to the presence of a precessing accretion disk in LMC\,X-2. This is
  strengthened by He\,II $\lambda$4686 Doppler maps that show a bright
  spot that is moving from night to night.  Furthermore, we detect
  narrow emission lines in the Bowen region of LMC\,X-2,with a
  velocity of $K_{\rm em}$=351$\pm$28 km s$^{-1}$, that we tentatively
  interpret as coming from the irradiated side of the donor star.
  Since $K_{\rm em}$ must be smaller than $K_2$, this leads to the
  first upper-limit on the mass function of LMC\,X-2 of
  $f(M_1)$$\ge$0.86$M_\odot$ (95\% confidence), and the first
  constraints on its system parameters.
\end{abstract}


\section{Introduction}

Low mass X-ray binaries (LMXBs) are compact binaries where the primary
is a compact object and the secondary a low mass star
($\le$1$M_\odot$). The secondary is transferring mass via Roche-lobe
overflow, forming an accretion disk around the compact object that
gives rise to the observed X-rays. By far, most of the persistently
bright LMXBs are neutron star systems that can be divided into two
groups, the Z-sources and Atoll sources (Hasinger \& van der Klis
1989). Z-sources are usually the brightest LMXBs in X-rays (they are
thought to have mass accretion rates that reach the Eddington limit)
and trace a Z-like shape in their X-ray colour-colour diagrams. Atoll
sources on the other hand have lower accretion rates ($\simeq$1-2
orders of magnitude lower) and their colour-colour diagrams usually
consists of fragmented 'island-like` regions. Apart from the
difference in accretion rates, the main physical difference between
Z-sources and Atoll sources are thought to be the strength of the
neutron star magnetic field and their evolutionary history (Hasinger
\& van der Klis 1989).

LMC\,X-2 is a persistent LMXB that shows the properties of a Z-source
(Smale et~al. 2003), and it is therefore thought to be a neutron star
system that has an accretion rate around the Eddington limit. It is
one of the most X-ray luminous LMXBs known ($L_X$$\simeq$10$^{38}$ erg
s$^{-1}$), but due to its extra-galactic nature (it is located in the
Large Magellanic Cloud at a distance of $\simeq$48 kpc), its X-ray
flux is rather low. Its optical counterpart was identified by Pakull
(1978) as a $B$$\sim$18.5 blue star. Despite being X-ray luminous and
having a known optical counterpart, thus far little is known about the
system parameters of LMC\,X-2. Even the estimates for the orbital
period range from 6.4 hrs (Motch et~al. 1985) or 8.2 hrs (Callanan
et~al. 1990, Smale \& Kuulkers 2000) up to 12 days by Crampton et~al.
(1990). To make things even more complicated, no periodic variability
was detected in 6 years of MACHO data (Alcock et~al. 2000).

\begin{figure*}
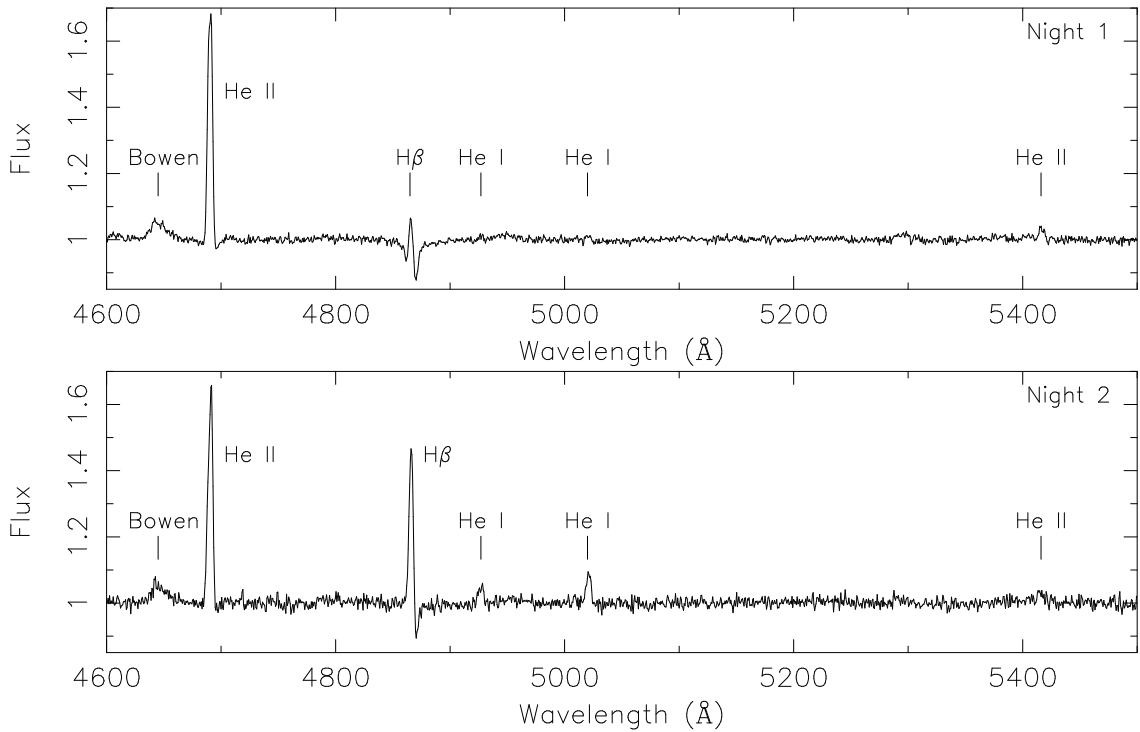

\psfig{figure=lmc_spec_night1.ps,angle=-90,width=15cm}
\psfig{figure=lmc_spec_night2.ps,angle=-90,width=15cm}
\caption{Average spectrum of LMC\,X-2 during the first observing night (top)
and the second night (bottom).
\label{spectrum}}
\end{figure*}

In recent years Steeghs \& Casares (2002) developed a new technique to
detect a signature of the donor star in persistent LMXBs. Using
phase-resolved spectroscopy they detected narrow emission lines in
Sco\,X-1, especially in the Bowen blend (a blend of N\,III and C\,III
lines between 4630-4650 \AA), that were interpreted as coming from the
irradiated side of the donor star. This discovery in Sco\,X-1 was
followed by a survey of other LMXBs that are optically bright enough
to also resolve these narrow components. Thus far these narrow
emission lines have been detected in X\,1822$-$371 (Casares et~al.
2003), GX\,339$-$4 (Hynes et~al. 2003), V801\,Ara and V926\,Sco
(Casares et~al. 2006), GR\,Mus (Barnes et~al. 2007), Aql\,X-1 and
GX\,9$+$9 (Cornelisse et~al 2007a,b), leading to constraints on their
system parameters.
 
In this paper we apply the technique of Bowen fluorescence to
LMC\,X-2.  We will show that it is possible to detect a periodic
signal in our spectroscopic dataset that we identify as the orbital
period. Furthermore, similar to the other X-ray binaries thus far, the
Bowen region shows the presence of narrow emission lines that we
identify as coming from the irradiated side of the companion, giving
the first ever constraints on the system parameters of LMC\,X-2.

\section{Observations and Data Reduction}

On November 21 and 22 2004 we obtained a total of 77 spectra of
LMC\,X-2 with an integration time of 600\,s each, using the FORS\,2
spectrograph attached to the VLT Unit 4 (Yepun Telescope) at Paranal
Observatory (ESO). Each spectrum was taken with the 1400V
volume-phased holographic grism using a slit width of 0.7$''$, giving
a wavelength coverage of $\lambda$$\lambda$4514-5815 and a resolution
of 70 km s$^{-1}$ (FWHM). The seeing during the first night was
between 0.4 and 0.7 arcsec, while on the second night it varied
between 0.5 and 2.7 arcs. The slit was orientated at a position angle
of 7$^{\circ}$ to include a comparison star in order to correct for
slit losses. During daytime He, Ne, Hg and Cd arc lamp exposures were
taken for the wavelength calibration scale. We de-biased and
flat-fielded all the images and used optimal extraction techniques to
maximise the signal-to-noise ratio of the extracted spectra (Horne
1986).  We determined the pixel-to-wavelength scale using a 4th order
polynomial fit to 20 reference lines giving a dispersion of 0.64
\AA\,pixel$^{-1}$ and rms scatter $<$0.05 \AA. We also corrected for
any velocity drifts due to instrumental flexure by cross-correlating
the sky spectra. Finally, we divided all spectra of LMC\,X-2 by a
corresponding low order spline-fit of the comparison star to get the
final fluxed spectra. Since we did not observe a spectro-photometric
standard star, we were not able to correct for instrumental response,
and all spectra are therefore in relative fluxes.

\section{Data analysis}

\subsection{Spectral characteristics}

We created average spectra for each individual night. Since we do not
have a flux standard to derive an absolute flux for LMC\,X-2, we
decided to normalise the continuum flux to one by dividing each
average spectrum by a low order spline fit. In Fig.\,\ref{spectrum} we
show the results. Both spectra are dominated by the very narrow high
excitation He\,II $\lambda$4686 emission line, while also Bowen
emission is present in both spectra. However, compared to other X-ray
binaries, such as Sco\,X-1, X\,1822$-$371, V801\,Ara and V926\,Sco the
Bowen emission is much weaker compared to He\,II in LMC\,X-2 (Steeghs \&
Casares 2002, Casares et~al. 2003, Casares et~al. 2006). This might be
due to the much lower metal abundances in the Large Magellanic Cloud
(Motch \& Pakull 1979).  The most striking difference between the
spectra is the dramatic change of H$\beta$ (see Fig.\,\ref{spectrum}).
During the first night (21 November) H$\beta$ is dominated by a weak
emission feature superposed on a broad absorption feature, while in
the second night (22 November) the emission feature has become almost
as strong as He\,II $\lambda$4686.  Furthermore, also the He\,I
$\lambda$4922$/$5016 lines have become more prominent during the
second night (although they might be present during the first night).

In order to quantify the change in the most prominent emission lines
we estimated the equivalent widths and their line fluxes (in arbitrary
units) for the two nights, and show them in Table\,\ref{width}. For
H$\beta$ we decided to also include the absorption component, giving
negative values for the first night.  Table\,\ref{width} shows that
there is no change in equivalent width of He\,II and the Bowen region,
and the line fluxes have dropped by $\simeq$40\% during the second
night. On the other hand the H$\beta$ and He\,I lines have all
increased significantly in both equivalent width and line flux.
Unfortunately, due to the faintness of these lines (or the presence of
an absorption feature), it is not possible to say if they have all
changed by the same amount, but it is likely that the same process is
responsible for this change in line intensity.  Finally, in order to
see if these changes could be related to a change in brightness we
have also created a lightcurve of the continuum flux for LMC\,X-2. We
have normalised the flux of the first night around unity, and show the
result in Fig.\,\ref{light}.  We note that during the second night the
continuum flux was $\simeq$40\% lower compared to the first night, a
similar fraction as was observed for the line fluxes of the He\,II lines
and Bowen region, keeping their equivalent widths the same and
suggesting a common origin.

\subsection{Radial Velocities}

We determined phase-resolved radial velocities by cross-correlating
each spectrum with a Gaussian of width 150 km s$^{-1}$
centered on the core of the He\,II $\lambda$4686 line. Since the
conditions during the second night were much worse, together with the
fact that the source was $\simeq$40\% fainter, we have binned these
spectra together in groups of three, and then determined the radial
velocity.  In Fig.\,\ref{radial} we show the results.

The first thing to note in Fig.\,\ref{radial} is that during both
nights the radial velocity shows a sine-like variation, which we
interpret as orbital motion of a region that is co-rotating with the
binary, and perhaps is connected to the dynamical properties of the
primary. However, during the second night it appears that either the
semi-amplitude of the radial velocity or the off-set compared to the
rest wavelength has increased. This last behaviour was noted by
Crampton et~al. (1990), who observed a long-term variation in the
radial velocity of He\,II.

We searched the radial velocity curve for any periodic signal with a
duration between 1 hr and 2 days using the Lomb-Scargle technique
(Scargle 1982). Apart from the 24 hr alias due to the separation of
our 2 observing nights, only two significant peaks with comparable
strength were present in the power spectrum; one peak is at a period
of 0.32$\pm$0.02 days and another at 0.45$\pm$0.05 day. Interestingly,
the 0.32 day period is similar to the photometric period detected by
Callanan et~al. (1990), suggesting that this is the orbital period.
Although we cannot exclude the possibility that the 0.45 day period is
real (but see Sect.\,3.3), we tentatively interpret the 0.32 day
period as the orbital period and use it in the rest of this paper.
Fitting a sine curve with a period of 0.32 day to the radial velocity
curve gives a phase zero at HJD 2,453,330.41$\pm$0.03, an off-set of
344$\pm$11 km s$^{-1}$, and the semi-amplitude of our sine fit is
41$\pm$4 km s$^{-1}$. Note that we did not attempt to account for the
seemingly variable systemic velocity from night to night, but only did
a single sine fit to both nights.

\begin{table}
\caption{LMC X-2 equivalent widths and spectral line fluxes.
\label{width}}
\begin{tabular}{lcccc}
\hline
line & Night 1  &       & Night 2  & \\ 
     & EW (\AA) & Flux  & EW (\AA) & Flux\\
\hline 
He\,II $\lambda$4686 & 3.20$\pm$0.06 & 182.7$\pm$0.9 & 3.07$\pm$0.08 & 112.1$\pm$1.2\\ 
Bowen                & 0.85$\pm$0.06 & 35.6$\pm$0.7  & 0.81$\pm$0.08 & 23.6$\pm$1.0 \\
He\,II $\lambda$5411 & 0.19$\pm$0.03 & 9.2$\pm$0.8   & 0.22$\pm$0.04 & 8.4$\pm$0.9  \\
H$\beta$             & -0.70$\pm$0.03& -43.0$\pm$0.9 & 1.59$\pm$0.06 & 49.9$\pm$1.1 \\
He\,I $\lambda$4922  & 0.04$\pm$0.03 & 3.8$\pm$1.1   & 0.26$\pm$0.04 & 15.1$\pm$1.4 \\
He\,I $\lambda$5016  & 0.03$\pm$0.03 & 3.3$\pm$1.1   & 0.44$\pm$0.04 & 27.5$\pm$1.4 \\
\hline
\end{tabular}\end{table}

\subsection{Doppler Maps}

We used Doppler tomography on the most prominent emission lines in
order to probe the structure of the accretion disk (Marsh \& Horne
1988). In order to create the maps, we used the orbital period of
8.16$\pm$0.02 hrs as determined by Callanan et~al. (1990) and a
systemic velocity derived from the radial velocity curve in
Sect.\,3.2. To comply with the standard definition of orbital phase 0
in Doppler maps (when the donor star is at inferior conjunction) we
used the phasing derived from the Bowen map (see below) and applied
a shift of $\simeq$0.5 orbital phase compared to the value derived
in Sect.\,3.2. 

Since He\,II $\lambda$4686 is by far the strongest emission line in
LMC\,X-2 we decided to create Doppler maps for each individual night,
and we show the result in Fig.\,\ref{hedopp}. Both maps show a
ring-like structure, but there are some minor differences. During the
second night it appears as if the outer edge of the structure is at
higher velocities than during the first night (218$\pm$10 km s$^{-1}$
compared to 180$\pm$10 km s$^{-1}$). Furthermore, during the first
night there is a clear emission feature in the lower part of the map,
that appears to have shifted toward the lower-right quadrant during
the second night.  Since we only have 2 nights of observations, it is
not clear if these changes are real, but it suggests that there is
some change in accretion disk structure from night to night.

We also created a Doppler map of the Bowen region by simultaneous
fitting all the major N\,III ($\lambda$4634/4640) and C\,III
($\lambda$4647/4650) lines using the relative strengths as given by
McClintock et~al. (1975). The map is dominated by a bright emission
feature, that we used to rotate the map (by $\simeq$0.5 orbital phase)
until it was located in the top. Fig.\,\ref{bowdopp} shows the
resulting map. The bright emission feature is at a velocity of $K_{\rm
  em}$=351$\pm$28 km s$^{-1}$, and another (much fainter) spot is also
present in the map. If we interpret the bright spot as arising on the
surface of the donor star (see Sect.\,4.2), the fainter spot is in a
region where we could expect an interaction between the accretion disk
and the accretion flow.  We do note that the velocity of the bright
spot is much higher than the disk velocities in the He\,II Doppler map
(Fig.\,\ref{hedopp}), and we will discuss this in Sect.\,4.2.
Furthermore, since the radial velocity curve in Sect.\,3.2 could not
exclude a 0.45 days orbital period, we also created a Bowen map for
this period. Although there are several spots present in this map,
none is as sharp and significant as those in Fig.\,\ref{bowdopp}. This
is further support that the 0.32 day period is the correct orbital
period.

\section{Discussion}

\subsection{Orbital period and the accretion disk}

We have presented phase-resolved spectroscopy of LMC\,X-2, one of the
brightest X-ray sources in the Large Magellanic Cloud, and also one of
the most luminous LMXBs known. This enables us to derive the first
constraints on its system parameters, and in particular give new
insights on the previously reported orbital periods that ranged from
$\simeq$6-8 hrs by Motch et~al.  (1985) or Callanan et~al. (1990) up
to $\simeq$12 days by Crampton et~al. (1990).

Callanan et~al. (1990) based their claim of a short orbital period on
an extended $\simeq$2 week photometric campaign. A clear 8.15 hr
modulation was present in their data that was interpreted as the
orbital period, although there is indication of a long term
($\simeq$10 day) variability that was also observed by Crampton
et~al. (1990). On the other hand, Crampton et~al. did not detect the
$\simeq$8 hr period in a $\simeq$1 week photometric campaign. However,
there was a variation in the He\,II radial velocity curve over a
period of 4 nights in their spectroscopic data that they interpreted
as a $\simeq$12 day orbital period. Interestingly, they noticed that
the H$\beta$ emission lines also changed in strength over those 4
nights (even going into absorption), with maximum line strength
occurring at minimum (continuum) light. Although Crampton et~al. (1990)
did speculate that the $\simeq$12 day period is a precession or beat
period, they discarded this due to the absence of shorter periods in
their data set.

\begin{figure}
\psfig{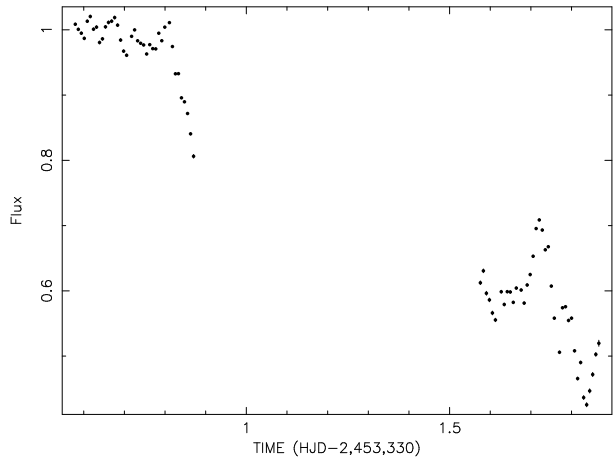}
\caption{Lightcurve of the continuum light of LMC\,X-2 where the first 
night is normalised to unity.
\label{light}}
\end{figure}

\begin{figure}
\psfig{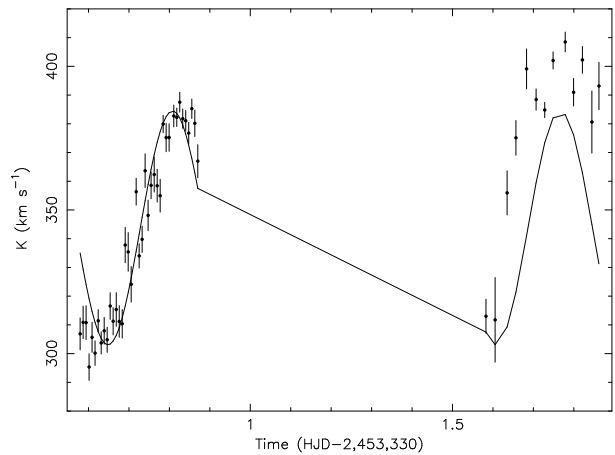}
\caption{Best fit radial velocity curve derived from the core of the 
He\,II $\lambda$4686 emission line. 
  \label{radial}}
\end{figure}

In Sect.\,3.2 we have shown that there is a $\simeq$8 hr period
present in the radial velocity of He\,II $\lambda$4686 that is similar
to the period detected by Callanan et~al. (1990). Since the
semi-amplitude of this radial velocity curve is rather small
($\simeq$41 km s$^{-1}$), it might have been difficult to detect with
the 4 m class telescope and an instrument with a resolution of $\sim$4
\AA~ used by Crampton et~al. (1990). We, therefore, tentatively identify
this period with the orbital period. However, that does leave the
question of the long term variation observed by both Crampton et~al.
(1990) and Callanan et~al. (1990). Similarly to Crampton et~al.
(1990), we also observe a large change in the emission line strength
of H$\beta$, with a stronger line occurring at lower continuum flux
levels.  Unfortunately we only have two nights of observations, and
are therefore not able to check if there is also a long term variation
in our He\,II radial velocity curve. However, Fig.\,\ref{radial} does
suggest that during the second night (when the continuum flux was
lower) either the amplitude or the mean velocity of the radial
velocity has increased compared to the first night (when the continuum
flux was higher), similar to what Crampton et~al. observed.

\begin{figure*}
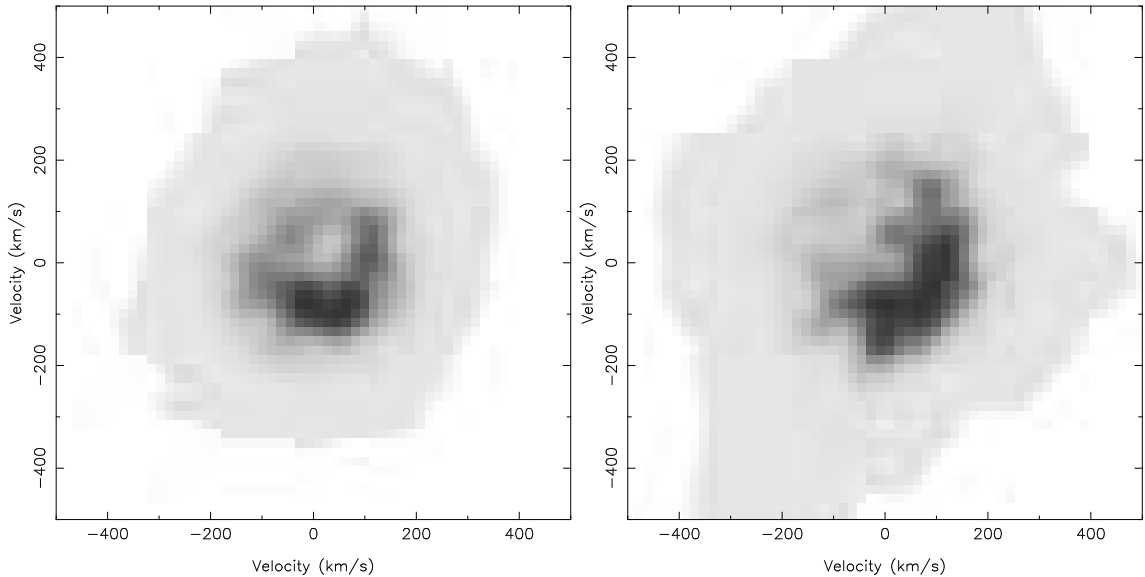

\parbox{5.8cm}{\psfig{figure=lmc_dop_heII1.ps,angle=-90,width=7.5cm}}
\parbox{5.8cm}{\psfig{figure=lmc_dop_heII2.ps,angle=-90,width=7.5cm}}
\caption{Doppler maps of the He\,II $\lambda$4686 emission line in LMC\,X-2 for
the first night of observations (left) and the second night (right). 
\label{hedopp}}
\end{figure*}

One explanation for the observed properties could be the presence of
an inclined precessing, warped, accretion disk in LMC\,X-2, as is also
observed in Hercules\,X-1 and SS\,433 (Katz 1973, Margon 1984). We will
discuss this suggestion in detail in a forthcoming paper by Shih
et~al. (in preparation), but here we will briefly highlight the
spectroscopic evidence. During the first night we could be observing
the accretion disk more edge on compared to the second night, and this
could explain the much lower H$\beta$ and He\,I line intensities
observed. This is further strengthened by the fact that the Doppler
maps suggest that He\,II is extending to higher radial velocities
during the second night. If true, this could also explain the change
in continuum flux observed in Fig.\,\ref{light}.  Callanan et~al.
(1990) already suggested that a significant contribution to the
optical light is either coming from the heated surface of the
secondary or the outer disk bulge. If the fraction of the secondary or
disk bulge that is in the shadow of the accretion disk changes as a
function of precession period, this would lead to a change in the
optical, with maximum light occurring when the accretion disk is most
edge on.

We can compare the characteristics of LMC\,X-2 to those of
XTE\,J1118$+$480, an X-ray transient that is known for having a
precessing (although not necessarily inclined) accretion disk (Uemura
et~al. 2000). Zurita et~al. (2002) showed that the nightly average
H$\alpha$ lines changed in both velocity and width that is consistent
with a periodic variation on the precession period (Torres et~al.
2004). Estimating the average wavelength of He\,II $\lambda$4686 for
the two nights in LMC\,X-2 gives 4690.37$\pm$0.01 and 4690.70$\pm$0.01
\AA, respectively, suggesting a slight velocity shift. However, we
must be careful with this slight shift, since we do not have a full
orbital coverage each night and this could lead to a systematic
off-set to the average wavelength. A better way to find out if this
slight shift in velocity is real, is by examining the two He\,II
Doppler maps. They show a bright spot that appears to have moved over
night, suggesting the presence of an irradiated region that shows
movement on a much longer time-scale than the orbital period, such as
the warped and irradiated part of the accretion disk.  Unfortunately,
we only have two nights of data and can therefore not follow the
long-term evolution of this bright spot to unambiguously claim that it
moves periodically on a longer timescale. A spectroscopic campaign with
a VLT-class telescope would be needed to follow the evolution of the
emission lines in LMC\,X-2 over a full expected precession cycle (of
$\simeq$1 week) and show that this bright spot in the He\,II Doppler
maps is long-lived and connected to a precessing disk.

\subsection{K-velocities?}

The radial velocity curve of the He\,II $\lambda$4686 emission line
shows a periodic variability that we have interpreted as the orbital
period. This could suggest that it also traces the primary and that we
have an estimate for both orbital phase zero and $K_1$. However, the
Doppler maps in Fig.\,\ref{hedopp} show that the He\,II emission is
dominated by the bright spot due to the warped accretion disk. Since
this spot does not have a similar phasing as the primary (and even
moves due to precession), we cannot use the radial velocity curve to
determine the orbital phasing of the primary. Furthermore, if LMC\,X-2
harbours an inclined precessing accretion disk it is also not likely
that semi-amplitude of the He\,II radial velocity curve traces $K_1$.
In this case the accretion disk, or at least the irradiated side
that produced the bright spot in the He\,II Doppler map, is tilted out
of the orbital plane thereby changing its radial velocity. This is
clear from Fig.\,\ref{radial}, where it appears that either the
average velocity or the semi-amplitude of the radial velocity curve
has changed. Only from long-term spectroscopic monitoring of LMC\,X-2
might it be possible to determine the $K_1$ velocity, but currently we
cannot constrain this value. This also means that currently we cannot
be certain that the average velocity that we determined corresponds to
the systemic velocity $\gamma$.

We have detected narrow emission lines in the Bowen region that
dominate the Bowen Doppler map. Although these lines are not visible
in the individual spectra, they become prominent when we create an
average spectrum that is shifted into the rest-frame of these narrow
lines. As Fig.\,\ref{average} shows, all important Bowen lines are
present, and especially the N\,III $\lambda$4640 line is very strong,
suggesting that this spot is real and not just a noise feature in the
Doppler map.  These narrow lines have been detected in many other
X-ray binaries thus far, such as Sco\,X-1, X\,1822$-$371, GX\,339$-$4,
V801\,Ara, V926\,Sco, Aql\,X-1, GX\,9$+$9 and GR\,Mus (Steeghs \&
Casares 2002, Casares et~al.  2003, Hynes et~al. 2003 , Casares et~al.
2006, Cornelisse et~al. 2007a,b, Barnes et~al. 2007). Since there are
few compact regions that could produce such narrow emission lines, it
was proposed that they arise on the irradiated surface of the donor
star.  Especially in X\,1822$-$371, but also in V801\,Ara this
connection could unambiguously be made, strengthening the claim in all
other sources.  Furthermore, the width of these emission lines in
LMC\,X-2 suggests that they come from a very compact region in the
system, and apart from the donor star surface not many other regions
in the binary could produce such narrow lines.  Therefore, following
the other systems and given the narrowness of these emission lines we
tentatively identify them as coming from the donor star of LMC\,X-2,
despite the fact that the absolute phasing of the system is unknown.

In LMC\,X-2 there is another problem with identifying the compact spot
in the Bowen Doppler map with the secondary, namely the fact that all
emission in the He\,II Doppler maps is at much lower velocities than
the compact Bowen spot. This would suggest that all emission in the
He\,II map is at sub-Keplerian velocities, and not related to the
accretion disk. However, such behaviour is not unique to LMC\,X-2.
Also in the He\,II $\lambda$4686 Doppler maps of many other LMXBs
(such as Sco\,X-1, X\,1822$-$371, V801\,Ara and V926\,Sco amongst
others) is most, if not all, emission at sub-Keplerian velocities
(Steeghs \& Casares 2002, Casares et~al. 2003, 2006).  However, if the
compact spot in the Bowen map is produced on the donor star, LMC\,X-2
would be the most extreme system in showing too low disk velocities.
We do note that one of the assumptions of the technique of Doppler
tomography is that all motion occurs in the orbital plane. If the
accretion disk in LMC\,X-2 is inclined, as suggested above, these
assumptions are no longer fulfilled. This would make the
interpretation of the accretion disk structure using the Doppler maps
more difficult, and could be a reason why we observe these
sub-Keplerian velocities. For example, in SS\,433 the precession angle
of the jets (and most likely also the accretion disk) is thought to be
$\simeq$20$^\circ$ (Margon 1984), and if LMC\,X-2 has a similar
precession angle (and a relatively low inclination) this could easily
account for observed accretion disk velocities that are a factor 2 too
low.  Again, long term spectroscopic monitoring would be needed to
give more insight into the real accretion disk velocities.  However,
we do note that this should not impact our interpretation of the
compact spot arising on the secondary star, and we remain confident
that we have detected a signature of the donor star.

\begin{figure}
\psfig{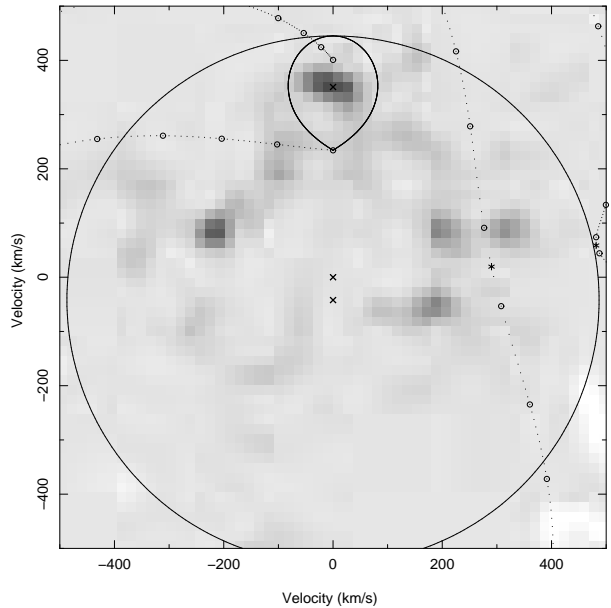}
\caption{Doppler map of the Bowen region in LMC\,X-2. The constrained spot
in the top of the diagram is at a position where a donor star signature is
expected. Superposed are the Roche-lobe,gas stream leaving the L1 point, 
and the Keplerian velocity along the stream for a random set of allowed 
parameters ($q$=0.12 and $K_2$=351 km s$^{-1}$. The circle indicates the 
Keplerian velocity at the outer edge of the accretion disk. 
\label{bowdopp}}
\end{figure}

\subsection{System parameters}

Assuming that we have detected the donor star in LMC\,X-2 we can use
these data to constrain the system parameters. Firstly, the narrow
lines must arise on the irradiated surface, therefore the determined
$K_{\rm em}$ (351$\pm$28 km s$^{-1}$) must be a lower limit to the
center of mass velocity of the secondary ($K_2$). However, since
$K_{\rm em}$ must be smaller than $K_2$ this already gives a lower
limit to the mass function of
$f(M)$=$M_1$$\sin^3$$i$/(1+$q$)$^2$$\ge$0.86 $M_\odot$ (at 95\%
confidence), where $q$ is the binary mass ratio $M_2$/$M_1$ and $i$
the inclination of the system. In order to further constrain the mass
function the $K$-correction must be determined.  Unfortunately, this
depends on $q$, $i$ and the disk flaring angle $\alpha$, all of which
are unknown for LMC\,X-2 (Mu\~noz-Darias et~al.  2005).  Furthermore,
the $K$-correction by Mu\~noz-Darias et~al.  (2005) assumes an
idealised accretion disk that is located in the orbital plane of the
system, both of which are most likely not true for LMC\,X-2.
Therefore, we must be very careful with any constraints we will derive
in the rest of this section.

\begin{figure}
\psfig{figure=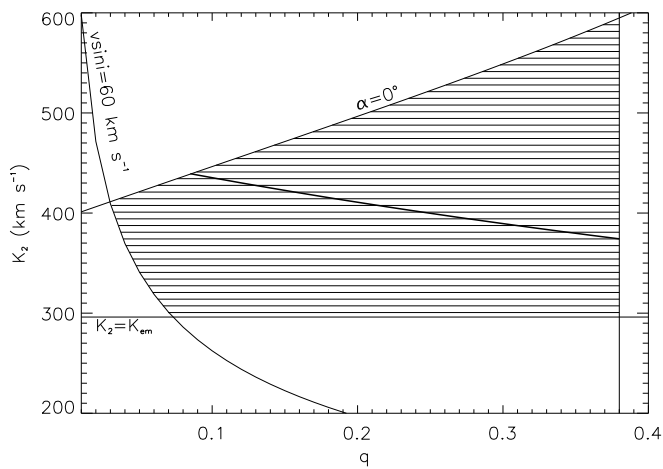,width=9cm}
\caption{Constraints on $q$ and $K_2$ for LMC\,X-2 indicated by the shaded
area. $K_2$ must be larger than $K_{\rm em}$, while the disk flaring angle must 
be larger than $\alpha$=0$^\circ$. Rotational broadening of the narrow lines 
gives the lower constraints on $q$ while the assumption that the accretion 
disk is precessing gives the upper-limit. The solid dark line drawn through the
shaded area indicates where $M_1$$\sin$$^3i$=3.2$M_\odot$; above this line the 
compact object must be a black hole.
\label{limits}}
\end{figure}

We can set a strict lower-limit to $K_2$ by assuming that
it is equal to $K_{\rm em}$, i.e that all emission is coming from the
poles.  Furthermore, we can use the 4th order polynomials for the
$K$-correction by Mu\~noz-Darias et~al. (2005) to determine $K_2$
as a function of $q$ in the case of $\alpha$=0$^\circ$ and
$i$=40$^\circ$ (since LMC\,X-2 does not show dips or eclipses its
inclination must be lower than $\simeq$70$^\circ$, and therefore the
polynomials for the $K$-correction when $i$=40$^\circ$ are a better
approximation than in the case of $i$=90$^\circ$). We show these limits
on $K_2$ in Fig.\,\ref{limits}. Note that these limits are still true even
if the accretion disk is severely warped or out of the orbital plane.

\begin{figure}
\psfig{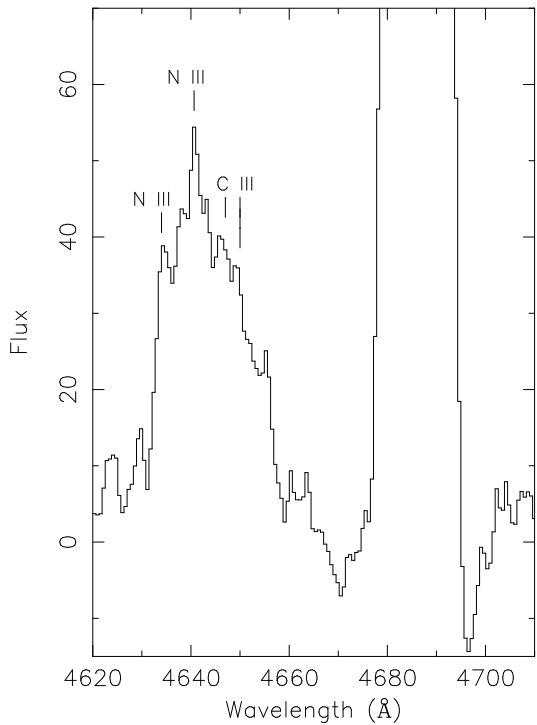}
\caption{Average spectrum of the Bowen region in the rest-frame of the narrow
emission lines in LMC\,X-2. We have indicated the main lines in this region.
\label{average}}
\end{figure}

Assuming that the width of the narrow emission lines is mainly due to
rotational broadening we can derive a lower-limit to $q$ using $V_{\rm
  rot}$$\sin$$i$=0.462$K_2$$q^{1/3}$(1+$q$)$^{2/3}$ (Wade \& Horne
1988). From Fig.\,\ref{average} we derive a {\it FWHM} of the narrow
lines of 90.2$\pm$18.8 km s$^{-1}$. Since these emission lines are
expected to arise from only part of the secondary, this value must be
a strict lower-limit and the true {\it FWHM} must be higher.
Furthermore, our estimate still includes the effect of the intrinsic
instrumental resolution of 70 km s$^{-1}$.  Following Casares et~al.
(2006) we accounted for this effect by broadening a strong line in our
arc spectrum using a Gray rotational profile without limb darkening
(since the fluorescence lines occur in optically thin conditions)
until we reached the observed {\it FWHM} (Gray 1992). We found that a
rotational broadening of $V_{\rm rot}$$\sin$$i$$\ge$60$\pm$18 km
s$^{-1}$ reproduced our results.  Since this is a strict lower-limit
on the true rotational broadening this will also give a lower limit on
both $q$ and $K_2$ that is shown in Fig.\,\ref{limits}.

Casares et~al. (2006) derived an upper-limit to $q$ by assuming that
the secondary is the largest possible zero-age main sequence star
fitting in its Roche-lobe. However, LMC\,X-2 shows the X-ray
properties of a Z-source (Smale et~al. 2003). Z-sources trace a Z-like
shape in their X-ray colour-colour diagrams and are thought to have
more evolved secondaries (see e.g. Hasinger \& van der Klis 1989).
Although the evolved nature of the secondary in Z-sources is not
unambiguously confirmed, it is still reasonable that the assumption of
a zero-age main sequence donor is most likely not true for LMC\,X-2.
We therefore cannot use this assumption to derive an upper-limit on
the mass ratio, since the evolved donor star could be more massive
than a main sequence star (Schenker \& King 2002). However, we can use
an alternative way to set an upper-limit to $q$, namely using the fact
that LMC\,X-2 has a precessing accretion disk. One of the main
criteria to produce a precessing accretion disk is to have an extreme
mass ratio (e.g.  Whitehurst 1988). From an overview by Patterson
et~al. (2005) of compact binaries that have a precessing accretion
disk and for which the mass ratio is known, we can use the system with
the largest measured $q$ thus far and assume that it must be smaller
for LMC\,X-2.  This leads to a conservative upper-limit of
$q$$\le$0.38, and our final limit for the system parameters shown in
Fig.\,\ref{limits}.  All these limits constrain an area in
Fig.\,\ref{limits} that is still quite large, and at the moment it is
still not possible to constrain any of the system parameters tightly
enough in order to say anything about the component masses or the
evolution of LMC\,X-2. However, we can make two more assumptions,
although more speculative, for LMC\,X-2 to further constrain its
system parameters.

Given that LMC\,X-2 shows the X-ray properties of a Z-source (Smale
et~al. 2003), we can speculate that its compact object is a neutron
star. In this case the maximum mass for the compact object would be
$\simeq$3.2$M_\odot$, and we could use this to set tighter
upper-limits to $K_2$ than derived from the assumptions we have made
thus far. In Fig.\,\ref{limits} we have therefore also indicated the
maximum limit where the compact object can still be a neutron star,
i.e. where $M_1$$\sin$$^3i$=3.2$M_\odot$, as a solid dark line.
However since we cannot completely rule out the possibility that the
compact object in LMC\,X-2 is a low mass black hole, we have decided
not to use this limit to constrain $K_2$. Finally, we can speculate
that the semi-amplitude that we have derived from the He\,II
$\lambda$4686 radial velocity curve is $K_1$. In this case we set an
upper-limit to $q$=$K_1$/$K_{\rm em}$$\le$0.12. Note that we have used
this value for $q$ and $K_2$=$K_{\rm em}$ to draw the Roche lobe in
Fig.\,\ref{bowdopp} to illustrate that it encompasses the bright
spot.  However, this set of system parameters is just one possible
solution within the allowed region in Fig.\,\ref{limits} and we do not
imply that it represents the true or even preferred system parameters
of LMC\,X-2.

We note that this value of $q$=0.12 is similar to the mass ratios
determined from other X-ray binaries that show a precessing accretion
disk (O'Donoghue \& Charles 1996), suggesting that it might be close
to the true mass ratio of LMC\,X-2.  Also, if we assume that the
$\simeq$12 day variation observed by Crampton et~al. (1990) is the
precession period, and that the 8.2 hr photometric period is a
superhump period, we can estimate the fractional period excess of the
superhump over the orbital period $\epsilon$. We find that
$\epsilon$$\simeq$0.03 and with the $\epsilon$($q$)-$q$ relation by
Patterson et~al. (2005) we estimate that $q$$\simeq$0.14. This is again
close to the mass ratio derived above.  Although this would set tight
constraints to $q$, and makes a black hole nature for the compact
object very unlikely, we have argued in Sect.\,4.2 that we cannot make
this assumption and have therefore not plotted this in
Fig.\,\ref{limits}. 

Finally, we do want to point out that, despite the large range of
system parameters possible, LMC\,X-2 most likely harbours a neutron
star that is more massive than the canonical 1.4$M_\odot$. Since
LMC\,X-2 does not show eclipses or dips we can estimate an upper-limit
on the inclination of 65$^\circ$ (see e.g.  Paczynski 1974). Combined
with the lower limit on the mass function already gives
$f$($M_1$)/$\sin^3$$i$$\ge$1.16$M_\odot$. For the compact object to be
1.4$M_\odot$, this only leaves $q$$\le$0.10 (and $M_2$$\le$0.14).
Furthermore, already for $K_2$=309 km s$^{-1}$ (and
$i$$\le$65$^\circ$) the mass of the compact object will exceed
1.4M$_\odot$. This only leaves a very small corner in the bottom-left
part of our allowed system parameters, and makes LMC\,X-2 another
strong candidate to harbour a massive neutron star.

\section{Conclusions}

We have detected for the first time a spectroscopic period in LMC\,X-2
that is close to previously published photometric period of 8.15 hrs
by Callanan et~al. (1990). We interpret this as the orbital period,
while the 12 day period detected by Crampton et~al. (1990) is most
likely the super-orbital period due to an inclined precessing
accretion disk.  The signature of such a precessing accretion disk is
also present in our data, mainly in the He\,II Doppler maps, but also
in the form of a nightly change in semi-amplitude and/or mean velocity
of the radial velocity curve of He\,II $\lambda$4686. In a forthcoming
paper by Shih et~al.  (in prep.) we will explore the consequences of
such a precessing accretion disk in more detail.

The main result of our spectroscopic data-set is the detection of
narrow emission lines in the Bowen region. Following previous
detections of such lines in other LMXBs, we tentatively identify these
as arising from the surface of the secondary. This gives us for the
first time the possibility to derive the mass function of LMC\,X-2 and
constrain its system parameters.  Although they point toward a massive
neutron star, the constraints on the system parameters are currently
not very tight. However, this could change with a better determination
of both the spectroscopic and precession period during a long term
spectroscopic campaign to study the emission line kinematics. Using
the relation between the mass ratio $q$ and the fractional period
excess of the spectroscopic and the photometric period $\epsilon$
could give strong constraints on $q$ and thereby the other system
parameters. In particular this would be an excellent way to identify
the nature of the donor star to find out if Z-sources really have more
evolved secondaries.

\section*{Acknowledgements}
This work is based on data collected at the European Southern
Observatory Paranal, Chile (Obs.Id. 074.D-0657(A)). We acknowledge the
use of the MOLLY and DOPPLER software packages developed by T.R.
Marsh. RC acknowledges financial support from a European Union Marie
Curie Intra-European Fellowship (MEIFT-CT-2005-024685). JC
acknowledges support from the Spanish Ministry of Science and
Technology through project AYA2002-03570.  DS acknowledges a
Smithsonian Astrophysical Observatory Clay fellowship as well as
support from NASA through its Guest Observer program. DS
acknowledges a PPARC/STFC Advanced Fellowship.

\bsp

\label{lastpage}

\end{document}